\documentclass[iop,apjl]{emulateapj}

\shorttitle{Discovery of Tidal Tails in Pal 14}
\shortauthors{Sollima et al.}
\begin{document}
\title{Discovery of tidal tails around the distant globular 
cluster Palomar 14}
\author{A. Sollima\altaffilmark{1,2}, D.
Mart\'{\i}nez-Delgado\altaffilmark{3,1,2}, 
D. Valls-Gabaud\altaffilmark{4}, J. Pe{\~n}arrubia\altaffilmark{5}}
%
%
%
\altaffiltext{1}{Instituto de Astrof\'{\i}sica de Canarias, 
               E38205 San Cristobal de La Laguna, Spain}
\altaffiltext{2}{Departamento de Astrofísica, Universidad de La Laguna, E-38205 San
Cristobal de La Laguna, Spain}
\altaffiltext{3}{Max-Planck-Institut f\"ur Astronomie,
                K\"onigstuhl 17, 69117 Heidelberg, Germany}
\altaffiltext{4}{GEPI, CNRS UMR 8111, Observatoire de Paris,
                5 Place Jules Janssen, 92195 Meudon, France}
\altaffiltext{5}{Institute of Astronomy, University of Cambridge, Madingley Road,
Cambridge CB3 0HA, United Kingdom}
\begin{abstract}
We report the detection of a pair of degree-long tidal tails associated with the
globular cluster Palomar 14, using images obtained at the CFHT. 
We reveal a power-law departure from a King profile at large distances to the
cluster center. 
The density map constructed with the optimal matched filter technique shows 
a nearly symmetrical and elongated distribution of stars on both sides of the 
cluster, forming a S-shape characteristic of mass loss. 
This evidence may be the telltale signature of tidal stripping in action. 
This, together with its large
Galactocentric distance, imposes strong constraints on its orbit
and/or origin: 
{\it i)} it must follow an external orbit confined to the
peripheral region of the Galactic halo and/or {\it ii)} it formed
in a satellite galaxy later accreted by the Milky Way.

\end{abstract}
\section{Introduction}
Globular clusters (GCs) are one of the cornerstones for understanding of the
formation, structure and dynamics of the halo of the Milky Way. 
Their dynamical evolution is driven both by internal mechanismçs (such as stellar 
evolution, two-body relaxation, and binary heating) and by external effects 
induced by the Galactic force field which produces the heating of their stars 
by tidal shocks during disk passages and tidal stripping. Both sets of effects 
lead to a continuous loss of stars and to the eventual dissolution of the cluster.
The stripped stars are placed on orbits similar to that of the original cluster, 
forming tidal tails surrounding the parent cluster.

A first attempt to search for tidal tails around GCs was carried out by
Grillmair et al. (1995) by analyzing the spatial distribution of star counts 
in a dozen Galactic GCs. They found that the observed density profiles deviate 
from the prediction of a best-fit King model at the outermost radii and extend 
beyond the conventional limiting radius set by this model.
Similar analyses were also done by Leon et al. (2000) using 2MASS 
data and more recently by Chun et al. (2010) with optical images.
Although all these studies detected spatially distinct star count overdensities around many 
clusters, density fluctuations caused by distant galaxy clusters, variable 
foreground reddening or photographic plate inhomogeneities may
seriously contaminate the stars counts, yielding in some cases uncertain
 locations and shapes of putative tidal tails (Law et al. 2003).
Two remarkable exceptions are the discoveries of 
extended tidal tails around \object{Palomar 5} (Odenkirchen et al. 2001)
and \object{NGC 5466} (Belokurov et al. 2006a). In both cases, a pair of tidal tails 
extending several degrees on the sky have been detected with high 
statistical significance (Odenkirchen et al. 2003; Grillmair \& Dionatos 2006; Zou et al. 2009). 
The orientation of the detected tails in \object{NGC 5466} is also in good 
agreement with its orbit as derived from proper-motion data.

While the above studies have focused on nearby objects, the sample of GCs in the 
outer halo of the Milky Way has 
thus far been excluded from these
studies since {\it i)} their distances make difficult to reach the most
populated regions of the color-magnitude diagram (CMD) which give an optimal
contrast for the detection of the cluster population against the foreground 
Galactic contamination, and {\it ii)} the tidal force exerted by the Galaxy in these remote
regions of the halo is expected to be too weak to produce significant
distortions in the cluster's shape (Lee et al. 2006; Fellhauer \& Lin 2007).
The most distant clusters for which signs of tidal disruption have been 
detected are all located within $20<R_{GC}<40~Kpc$ (Cote et al. 2002; Carraro et al.
2007,2009; Niederste-Ostholt et al. 2010).
Nevertheless these remote clusters represent an important class of objects to 
investigate many topics related to the hierarchical build-up process of the Galactic 
halo (Prieto \& Gnedin 2008) as well as fundamental physics. 
Indeed, in the classical Searle \& Zinn (1978) scenario of the formation of
the Galaxy, at least part of the halo GCs formed in external dwarf galaxies later accreted by the Milky Way.
This hypothesis is supported by the evidence that the group of 8 GCs populating the outermost Galactic halo
(at Galactocentric distances $R_{GC}>40$kpc) does not show the clear metal abundance gradient 
observed in the inner parts of the Galaxy and exhibit peculiar kinematics 
(large, energetic orbits of high eccentricity), larger core radii and a higher
specific frequency of RR Lyrae stars (Mackey \& Gilmore 2004).
Moreover, because of the small acceleration produced by the Milky Way in these remote
regions, these clusters 
represent also an excellent benchmark to test the gravitational law in such regimes 
(Baumgardt et al. 2005; Sollima \& Nipoti 2010; K\"upper \& Kroupa 2010).

Here we report the detection of a tidal tail around  \object{Palomar 14}, a GC
located in the outer halo of the Milky Way at a distance 
of $\sim$ 71 kpc. 
This result comes from a photometric campaign performed at the CFHT to search for extra-tidal structures in the 
outskirts of Galactic GCs (see
Martinez-Delgado et al. 2004) and has deep implications on the nature of
this cluster and its orbit.

\section{Observations and Data reduction}

Wide-field photometric imaging was obtained in QSO mode with the MegaCam camera 
at the Canada-Franch-Hawaii Telescope (CFHT) in 3 different nights in April and May 
2009.
The camera consists of a mosaic of 36 chips with a pixel scale of 
0.185" pixel$^{-1}$ providing a global field of view of 
$\sim~1^{\circ} \times 1^{\circ}$. A set of 6 $g'$ and 9 $r'$ 680 sec-long exposures 
were taken around the cluster center with a dithering pattern of few arcminutes 
to fill in the gaps between the chips. The average seeing was 0.7".

The standard reduction steps (bias, dark and flat-field correction) were carried out with the Elixir pipeline developed by the CFHT team.
We used DAOPHOT II and the point-spread-function (PSF) fitting algorithm 
ALLSTAR (Stetson, 1987) to obtain instrumental magnitudes for all the stars 
detected in each frame. The TERAPIX pipeline was then used to produce
mean frames by aligning and averaging the  images
with a 3$\sigma$ clipping rejection threshold. 
The automatic detection of sources was performed on the mean frames adopting a 
3$\sigma$ threshold. The mask with the object positions was then used as input 
for the PSF fitting, which was performed independently on each image. 
The most isolated and bright stars in each field were used to
build the PSF model (here a Moffat function of exponent 2.5).
For each passband, the derived magnitudes were transformed to
the same instrumental scale and averaged. We adopted the nightly zero points 
and reddening coefficients provided by the CFHT to link the
instrumental magnitudes to the standard system.
Finally, a catalog with over 100,000 calibrated sources was
produced and astrometrically calibrated through a cross-correlation
with the DR7 release of the Sloan Digital Sky Survey (Abazajian et al. 2009), 
which lists accurate positions for some 40,000 objects over an area of 
$\pi$ sq.deg. around Palomar 14. The astrometric
solution has a typical standard deviation of 200 mas.

\section{Results}

\subsection{Color-Magnitude Diagrams}

\begin{figure}
\epsscale{1.2}
\plotone{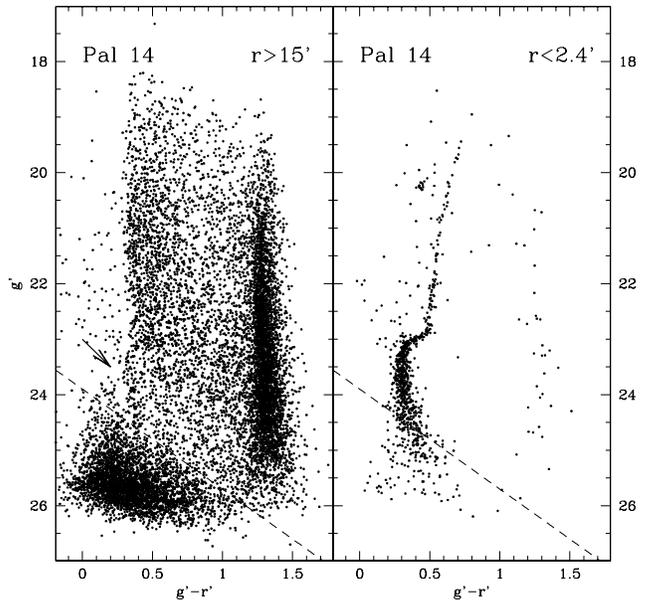}
\caption{CMDs of the surveyed area around Palomar 14. The right panel shows
the CMD of the innermost 2.4', the left panel shows the CMD of stars located at
distances $r>15'$ to the cluster center. Only stars with a sharpness parameter 
$\vert S \vert < 0.2$ are plotted. In the left panel the observed overdensity 
of stars in the outer field is indicated.
The dashed lines indicate the limit 
below whom the sharpness criterion do not provide a good star/galaxy discrimination.}
\label{cmd}
\end{figure}

The CMDs ($g',~g'-r'$) of the
innermost region of \object{Palomar 14} (within the half-light
radius $r_h \sim$ 2.4') and of the
most external portion within the Megacam field of view (at distances 
$r>$15' to the
cluster center) are shown in Fig.~\ref{cmd}. Only objects with a sharpness parameter $\vert S\vert<$0.2 (as
defined by Stetson 1987) have been plotted to minimize the contamination from
background galaxies. The CMDs sample the evolved population of the cluster, 
reaching the Main Sequence (MS) at $g'\sim 25$. 
A significant overdensity of stars can be noticed in the external region at
$g'-r'\sim 0.5$ and $23<g'<24.5$, with a
morphology and magnitude which resembles the MS of Palomar 14, at a distance much 
larger than the estimated tidal radius 
(between 7.1' and 9.2'; McLaughlin \& van der Marel 2005).

\subsection{Spatial distribution}
\label{s:spat}

\begin{table*}
\begin{center}
\caption{Best-fit parameters of the density profile of Palomar 14.}
\label{tbl-1}\begin{tabular}{lccccr}
\tableline\tableline
Model & $W_{0}$ & $r_{c}$ & $r_{h}$ & $r_{t}$ & $M_{V}$\\
      &         &  $'$    & $'$     &  $'$  &\\
\tableline
King    & 7.0 & 0.60 ($\pm$0.07) & 2.37 ($\pm$0.15) & 20.0 ($\pm$1.5)& -4.94 ($\pm$0.12) \\
Wilson  & 6.3 & 0.72 ($\pm$0.08) & 2.23 ($\pm$0.14) & 27.2 ($\pm$1.7)& -4.95 ($\pm$0.12) \\
Plummer &     & 2.21 ($\pm$0.30)\tablenotemark{a} & 3.91 ($\pm$0.53)& & -4.65 ($\pm$0.25) \\
\tableline
\end{tabular}
\tablenotetext{1}{For the Plummer model we report the characteristic radius.}
\tablecomments{From the isochrone fitting we derived a distance $d=71\pm2~Kpc$
and an age t=$13.2\pm0.3$ Gyr,
assuming a metallicity of [Fe/H]=-1.6 (Armandroff et al. 1992) and
[$\alpha$/Fe]=+0.3 (Ferraro et al. 1999).}
\end{center}
\end{table*}

\begin{figure}
\epsscale{1.2}
\plotone{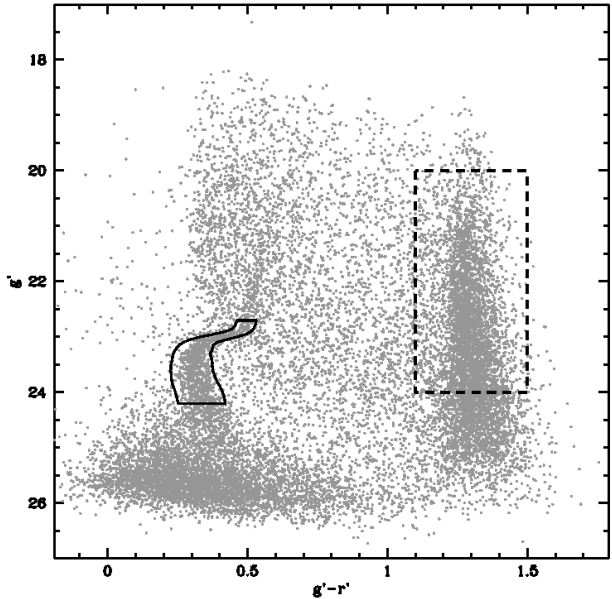}
\caption{CMD over the entire Megacam field of view centered on Palomar 14.
The selection boxes for the cluster (solid line) and  field 
(dashed line)  populations are indicated.}
\label{sel}
\end{figure}

\begin{figure}
\epsscale{1.2}
\plotone{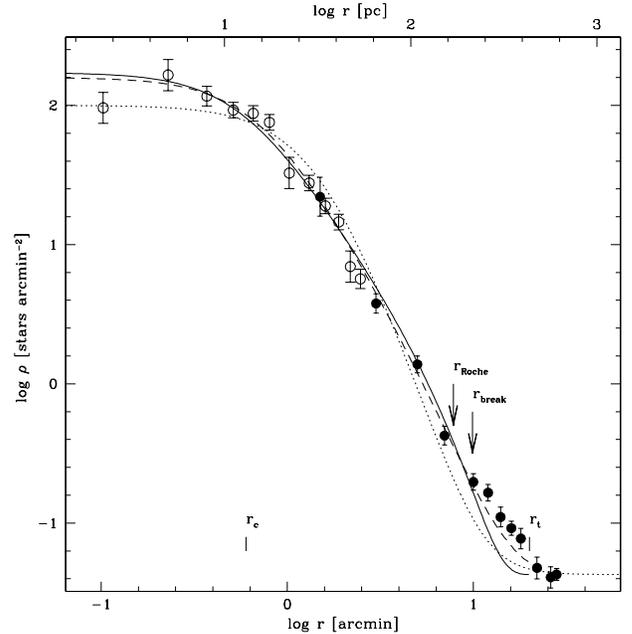}
\caption{Radial density profile of Palomar 14. Open circles are from Trager et al.
(1995), filled circles are from this work. The best fit King (1966; solid line), 
Wilson (1975; dashed line) and 
Plummer (1911; dotted line) models are indicated (see Table \ref{tbl-1}).
A constant background has been assumed (see text). The locations of the core and tidal
radii are indicated together with the maximum Roche
lobe and the break radii (vertical arrows).}
\label{rad}
\end{figure}

To study the spatial distribution of the stars of Palomar 14 
we first computed its radial density profile. 
For this purpose we selected those stars in the magnitude range 
$22.7<g'<24.2$ which lie within 3 times the local color dispersion about the 
cluster mean ridge line to ensure a good level of completeness and to minimize the contamination from 
Galactic field stars (see Fig.~\ref{sel}).
The number of selected stars in concentric annuli of variable
width located at various distances to the cluster center (from 1' to 28')
were counted to produce the surface density profile shown in Fig. \ref{rad}. 
In the innermost region (where the MS star counts suffer from a significant degree of
incompleteness) our profile has been integrated with surface brightness measures by Trager
et al. (1995) converted in densities and scaled to properly match our measures in the
overlapping regions.   
The radial profile extends well beyond the tidal
radius estimated so far (Harris \& van den Bergh 1984; Trager et al. 1995; McLaughlin \& van der
Marel 2005) and in fact the distribution appears to be truncated by the MegaCam 
field of view.
The best fit King (1966), Wilson (1975) and Plummer (1911) 
models are overplotted in Fig. \ref{rad}\footnote{In the model fitting a constant background of density 
$\log~\rho=-1.37~\mathrm{stars}~\mathrm{arcmin}^{-2}$ has been assumed from the MS star counts in
the most external region of our image (at $r>$25'). This value is in agreement
with the predictions of the Galactic model of Girardi et al. (2005).}.
In the fitting procedure we excluded those points located beyond $r>10'$.
In fact, in these outer regions the density profile deviates from the behavior predicted
by all models, declining with a power-law exponent
$\alpha=-1.6$, in agreement with the values found in other Galactic GCs 
(Grillmair et al. 1995; Leon et al. 2000; Testa et al. 2000; Lee et
al. 2003) and as predicted by theoretical models (e.g. Johnston et al. 1999).
Our best fit King and Wilson models predict 
respectively a tidal radius of 20.3'$\pm 1.7$
 and 27.2'$\pm$ 1.7 (i.e., four times larger than previous estimates) 
 while clearly underestimating the stellar density in the outermost radii.

\begin{figure}
\epsscale{1.2}
\plotone{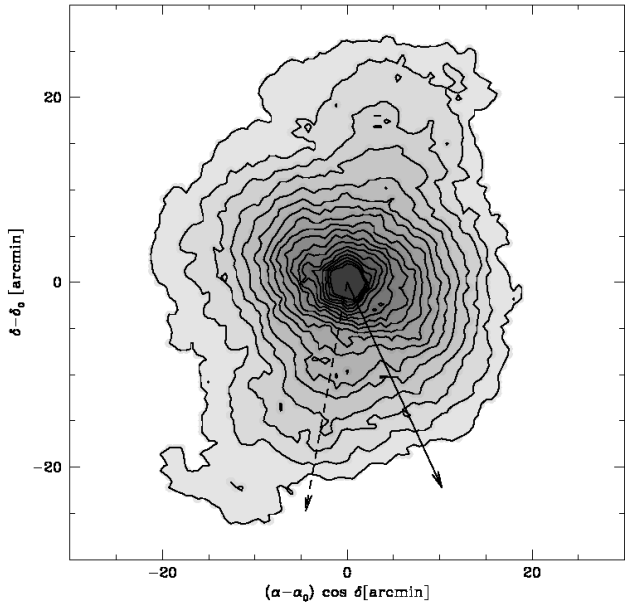}
\caption{Large-scale surface density around Palomar 14. The contour levels range from 3 
to 20$\sigma$ above the background
density in logarithmic steps of 1$\sigma$ each. The direction of the Galactic
center and of the predicted proper motion by Lynden-Bell \& Lynden-Bell (1995)
are indicated by the solid and dashed arrows, respectively.}
\label{map}
\end{figure}

To investigate in detail the two-dimensional  distribution of cluster stars we apply an
optimal matched filter technique (see Kuhn et al. 1996;
Rockosi et al. 2002; Odenkirchen et al. 2001, 2003).
Briefly, we define a fiducial cluster and field population in the CMD by 
sampling all stars that lie within 2.4' and 
outside 25'~ respectively\footnote{As shown in
Fig. \ref{rad}, Palomar 14 extends well beyond this limit and some cluster 
stars may fall inside
the reference field population. However, given the large difference in density between
the Galactic field population and the cluster one in this outer region, the
effect of this contamination on the derived weights is negligible.}. The densities in the CMDs (Hess diagram) of the fiducial
cluster and field populations were 
computed using an adaptive kernel estimation (Silverman 1986) with a Gaussian
kernel of radius set to the distance of the 10th nearest star. 
We then assigned to each star a weight defined as the ratio between the densities
calculated at that position of the cluster and the field CMDs. 
Lastly, the distribution of stellar  
positions was transformed into a smoothed surface density function through an 
adaptive kernel estimation. In this case we used a Gaussian kernel with
 radius set to the angular distance of the 100th nearest 
neighbor of each star and the kernel volume was set proportional to the associated 
weight. This procedure yields the surface density distribution shown in 
Fig. \ref{map}. As apparent,the overall distribution presents an elongated shape. In
particular, the density contours have a symmetrical distribution to both sides 
of the cluster with an orientation that appears to change with distance, forming
the characteristic S-shape typical of stellar mass loss. 
The direction of the Galactic center is also indicated in Fig. \ref{map}. It is
interesting to note that the elongation of the central region of the cluster is
not too different from this direction.

\begin{figure}
\epsscale{1.2}
\plotone{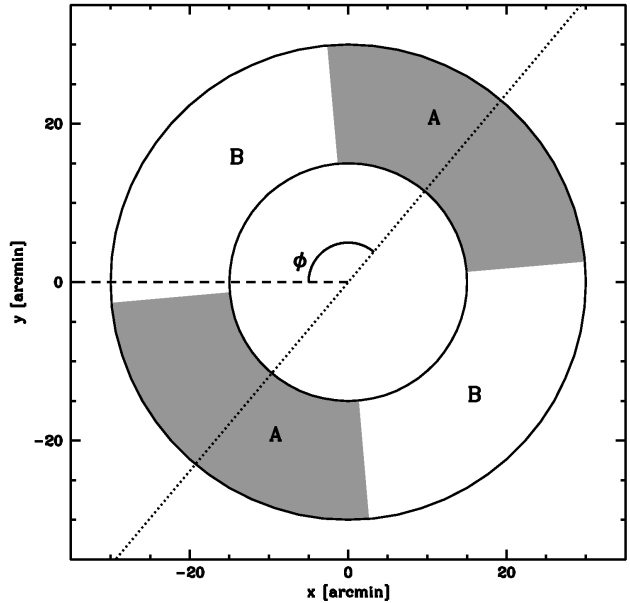}
\caption{Map of the selected areas for the coherence test. The A and B
regions for a given position angle $\phi$ are shown as grey and empty areas,
respectively.}
\label{vis}
\end{figure}

\begin{figure}
\epsscale{1.2}
\plotone{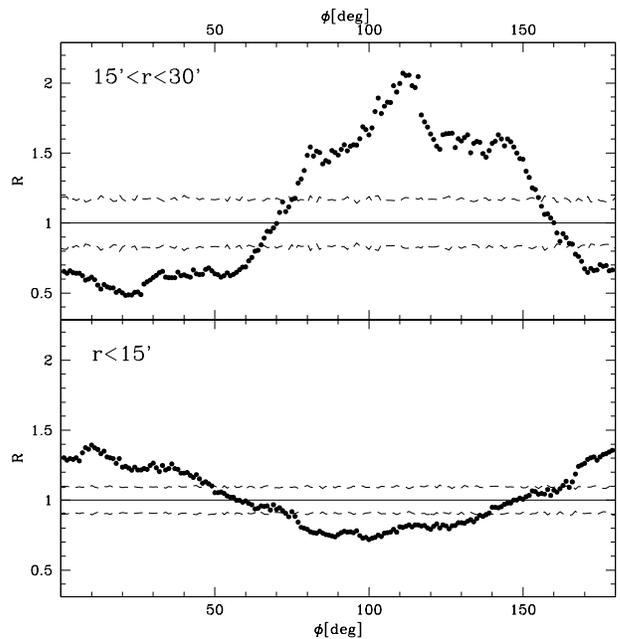}
\caption{The test statistic ${\cal R}$ as a function of the position angle $\phi$ 
 for stars at $r<$15' ({\it bottom panel}) and 15'$<r<$30' ({\it top
panel}). The mean (solid line) and standard deviation (dashed) of Monte Carlo samples of 
homogeneous distributions are also indicated to assess the statistical 
significance.}
\label{ang}
\end{figure}

\begin{figure}
\epsscale{1.2}
\plotone{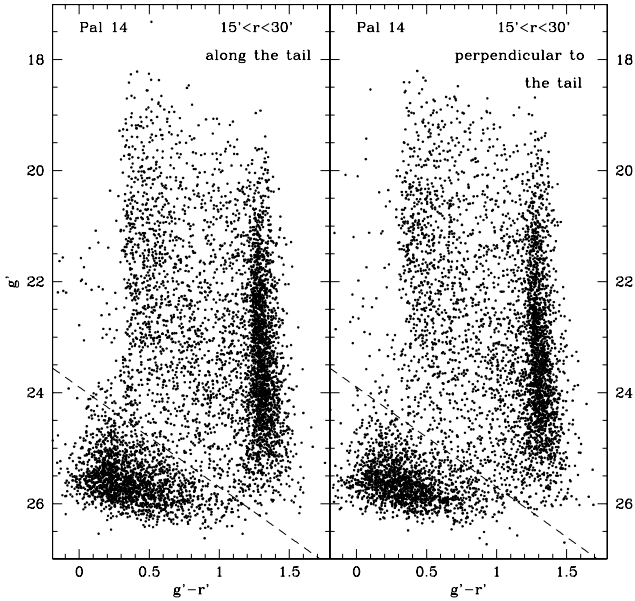}
\caption{Color-magnitude diagrams of  regions A (along the detected tails; 
{\it left panel}) and B
(perpendicular to the tails; {\it right panel}) at position angle
$\phi=111^{\circ}$~ for
stars at distances 15'$<r<$30'~ and sharpness $\vert S \vert<0.2$.}
\label{cmdt}
\end{figure}

As a further check of the anisotropic distribution of cluster stars, we 
can quantify the degree of coherence in the orientation of
the tails. We consider the sample of stars
selected on the CMD close to the cluster mean ridge line (as defined above) and a sample of M dwarfs of
the Galactic disk (selected in the range $1.1<g'-r'<1.5$ and $20<g'<24$;
see Fig. \ref{sel}).
This control sample is supposed to be uniformly distributed across the field of
view and actual deviations from the uniform distribution reflect variations in the efficiency of
detection, Galactic field and extinction gradients, etc.  
We define two regions (A and B) as alternate pairs of $90^{\circ}$-wide circular 
sectors positioned at a given distance to the cluster center and oriented at a
given position angle ($\phi$) in opposite directions (see Fig. \ref{vis}).
For a given position angle, we count the number of cluster ($N_{c}$) and 
field ($N_{f}$) stars lying in the regions A and B and compute the normalized
ratio
${\cal R}(\phi)=(N_{c}^{A} \, N_{f}^{B}) / (N_{f}^{A} \, N_{c}^{B})$. 
We performed this test using both stars 
located at distances $r<$15'~ and stars at 
$r>$15'.
The statistical significance of the test was evaluated with Monte Carlo
simulations to overcome to the notorious misbehavior of the ratio of Poisson
variables (e.g. Cervi\~no \& Valls-Gabaud 2003).
We constructed a set of 1000 simulations placing the same number of observed 
cluster and field stars at random angles and computed the ratio $\cal{R}$ for each
simulated set, yielding the expected trend of ${\cal R}$ and its standard deviation 
for a homogeneous and isotropic distribution of stars. The result of such a test is shown in Fig. \ref{ang}.
As expected, the
simulated samples yield a value of ${\cal R}$ close to unity for all
position angles. Instead, the observed ratio presents a clear peak in both areas. In
particular, the maximum coherence is found at the position angle of
$\phi=10^{\circ}$ (inner region) and  $\phi=111^{\circ}$ (outer region). In addition, 
the maximum value of ${\cal R}$ is larger in the outer
region (${\cal R}$=2.07; corresponding to a statistical significance of 6.1$\sigma$)
 than in the 
inner one (${\cal R}$=1.39; at 3.7$\sigma$). 
Fig. \ref{cmdt} shows 
the CMDs of the regions A (along the tail) and B (perpendicular to the tail) 
at distances $r>$15'. The density in the MS portion of the CMD is
significant along the tail, while only a sparse number of stars is
visible in the region perpendicular to the tails.

.
\subsection{Cluster Structure and Relaxation}
\label{struc}

The density profile shown in Fig. \ref{rad} is the deepest and most complete one
currently available for this cluster and allows to calculate its physical
size and mass. First, the distance modulus is measured using the Bayesian inference
method (Hernandez \& Valls-Gabaud 2008), assuming metallicity and alpha 
enhancement marginalizing over the "nuisance" 
parameters of extinction, yielding maxima in the
posterior probability distribution function of distance and age. We used the
theoretical isochrones by Marigo et al. (2008) specifically transformed into the
MegaCam filters photometric system and assumed [Fe/H]$=-$1.6 (Armandroff et al.
1992) and [$\alpha$/Fe]=+0.3 (Ferraro et al. 1999).
We obtained a best-fit distance of 71$\pm$ 2 kpc and an age of 13.2$\pm$0.3 Gyr,
 in agreement with the estimates by 
Jordi et al. (2009) ($d=71\pm$1.3 kpc and
t=11.5 Gyr, assuming [Fe/H]$=-$1.5 and [$\alpha$/Fe]=+0.2).
The best fit Wilson model gives a projected
half-light radius of $r_{h}$=2.23'$\pm 0.14'$=46.1 $\pm 2.9$~pc which is 
the largest among those listed by Harris (1996) for Galactic GCs.

We then calculated the mass of Palomar 14 adopting two different methods.
First, we assumed a M/L ratio of 1.885 
(as derived by McLaughlin \& van der Marel 2005 using suitable stellar population
synthesis models) and a luminosity of $log~(L/L_{\odot})=3.91\pm0.05$. 
This last quantity has been derived from the cluster absolute V magnitude 
($M_{V}=-4.95\pm0.12$) calculated 
by integrating the bestfit surface brightness Wilson model 
and assuming the distance modulus $(m-M)_{V}=19.41\pm0.12$ calculated above.
The overall mass of the cluster turns out to be $log~(M/M_{\odot})=4.19\pm0.05$.
A second estimate has been done by using the velocity dispersion estimated by 
Jordi et al. (2009; $\sigma_{v}=0.38\pm0.12~Km~s^{-1}$), the half-light radius
calculated above and adopting eq. 4
of Baumgardt et al. (2005), yielding a value of $log~(M/M_{\odot})=4.14\pm0.14$. 
The two above estimates agree quite well between them.

It is also possible to estimate the time elapsed since the last pericentric 
passage using the observed velocity dispersion 
by Jordi et al.(2009) and the position of the break radius in the radial profile 
($r_{break}$; defined as the point of intersection between the best-fit
Wilson model and the external power-law profiles). 
In fact, using eq. 5 of Pe{\~n}arrubia et al. (2009) and the above value of
$r_\mathrm{break}$, we get $(t-t_{p})=1.0\pm0.3$~Gyr. The stars that populate 
now the tails are supposed to be mainly those stripped in the
last orbit. In this sense, the above estimate can be considered as the age of
the tail.

Adopting the Galactic
potential of Johnston et al. (1995) and the mass $log(M/M_{\odot})=4.19$ calculated
above, the maximum radius of the Roche lobe of a 
stellar system located at the position of Palomar 14 and with a projected radial velocity
of 72.3 km~s$^{-1}$ (Jordi et al. 2009) turns out to be 170 $\pm 10$ pc (see eq. A2 of 
Allen et al. 2006). This translates into an angular distance of
 $r_\mathrm{Roche}=8.2'\pm 0.4$. It is interesting to note that this 
 limit is located
well within the cluster tidal radius (see Fig. \ref{rad}) indicating that 
a number of cluster stars within the tidal radius are going to be lost in the
next pericentric passages. 

An observational estimate of the undergoing destruction rate in Palomar 14 
can be obtained (assuming that light traces mass) by counting the number 
of stars ($n_{b}$) comprised within
the break radius and those ($n_{xt}$) in the radial range 
$r_{break}<r<r_{t}$  as
$$\nu=-\frac{1}{M}\frac{dM}{dt}=\frac{r_{break}}{r_{t}-r_{break}}
\frac{n_{xt}}{n_{break}}\frac{\pi}{P_{orb}~cos \theta}$$
where $\theta$ is the angle between the line of sight and the plane
perpendicular to
the orbit, and $P_{orb}$ is the orbital period (Johnston et al. 1999).
While these two parameters depend on the unknown orbit of the cluster,
a lower limit to the rate can be derived assuming a
circular orbit 
($\cos\theta\sim 1, \, P_{orb}=2\pi R_{GC}/v_{circ} = 1.5\,(R_{GC}/50\,\mathrm{kpc}) \, Gyr \sim 2.2\,Gyr$). 
Given $r_{b}=9.9'\pm0.6'$, $r_{t}=27.2'\pm1.7'$, we
measure  $n_{xt}/n_{break}=0.21\pm0.02$  
and thus infer a lower limit to the destruction rate of $\nu > 0.18\pm0.06\,Gyr^{-1}$.

An important question arise regarding the mechanism which drives the formation
of tidal tails in this cluster in spite of its large Galactocentric distance. In
particular, it is worth discussing the importance of relaxation
and tidal effects in the process of mass loss of this cluster.

Remarkably,  
the half-mass relaxation time, defined as
$$t_{rh}=\frac{0.138}{<m>~ln(0.4~M/<m>)}\left(\frac{M~r_{rh}^{3}}{G}\right)^{1/2}
$$ 
\hspace{5cm} Spitzer \& Hart (1971)\\ \\
turns out to be
$t_{rh}=19.9\pm0.9$ Gyr. This quantity is longer than the age of the
Universe and indicates that relaxation should have produced small dynamical 
effect in the recent history of Palomar 14.
In the above calculation, we assumed a mean stellar mass of
$<m>=0.42~M_{\odot}$ and the mass $log~(M/M_{\odot})=4.19\pm0.05$ estimated
above.
It is interesting to calculate the change in the estimated relaxation time 
due to a different assumption of the cluster mass function.
In particular, we considered the case of a mass function equal to the Initial Mass
Function by Kroupa (2001; which is the case for a non-relaxed system) and the
mass function proposed by De Marchi et al. (2005; depleted in the low-mass
range thus simulating the effect of mass loss in a relaxed system). 
By using the Marigo et al. (2008) isochrones and the initial-final mass relation
by Kruijssen (2009) we estimated M/L=1.636,
$<m>=0.31~M_{\odot}$ and $t_{r}=24.5~Gyr$ for the Kroupa (2001) mass function 
and M/L=1.351, $<m>=0.42~M_{\odot}$ and $t_{r}=17.4~Gyr$ for that of De Marchi et al. (2005)\footnote{Note that the
M/L ratios calculated here are smaller than that provided by McLaughlin \& van der
Marel (2005) because of the different stellar models adopted by these authors.
This difference increases furthermore the estimated relaxation time, reinforcing 
the derived conclusion.}.    
Therefore, although the adoption of a different mass function can reduce the
estimated relaxation time of $\sim30\%$, it always results larger than the
cluster age.
According to the above considerations, it seems that relaxation cannot be
responsible for the recent mass loss experienced by Palomar 14.
Note however that relaxation driven expansion causes clusters to evolve towards 
larger half-mass radii as mass loss proceeds (Gieles et al. 2010), increasing the estimated relaxation time. It is therefore possible
that during its past evolution the relaxation time of Palomar 14 was
shorter than its age. In this regard, the lack of relaxation in Palomar 14 is 
strongly supported by the non-segregated radial distribution of the massive 
Blue Straggler Stars (Beccari et al., in preparation).

\begin{figure}
\epsscale{1.2}
\plotone{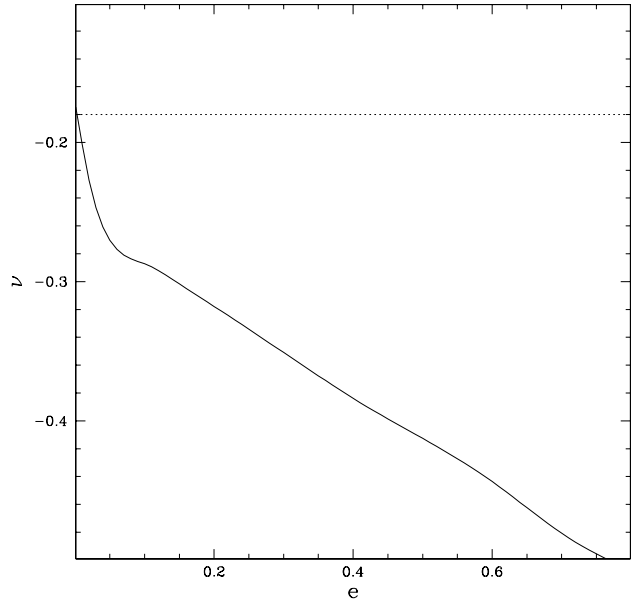}
\caption{Last orbit destruction rate as a function of the orbital eccentricity calculated
in the case of tidal stripping from a non-relaxed satellite 
(solid line; calculated from Taylor \& Babul 2001). 
The observational lower limit of the current destruction rate is 
marked with a dotted line.}
\label{mdot}
\end{figure}

Another important mechanism of mass loss is due to the tidal interaction of
Palomar 14 with the Galactic potential.
In this case the mass loss is due to the continuous stripping of "former" 
extra-tidal stars (i.e. the same process at work in the 
tidal destruction of Sagittarius) and by tidal shocks. 
The evidence that a significant fraction of cluster stars exceed the
present-day tidal radius suggests that this effect can be significant in this 
stellar system.    
Taylor \& Babul (2001) performed
extensive N-Body simulation to estimate the mass loss rate for the above 
mechanism and provided useful analytical relations.
To test the efficiency of this mechanisms, we estimated the predicted 
destruction rates for different orbital eccentricities.
For this purpose, we assumed the cluster presently at apocenter and adopted the 
cluster structural parameters of the best-fit Wilson (1975) model reported above. 
We adopted the method outlined by Taylor \& Babul (2001), assuming the
cluster orbiting in the logarithmic Galactic potential defined by Baumgardt 
\& Makino (2003).
In Fig. \ref{mdot} the mass loss rate estimated for the last orbit of Palomar 14 
is shown as a function of the orbital eccentricity. 
It is apparent that tidal stripping is very efficent even at small orbital eccentricity. 
The observational estimate of the disruption rate is also indicated in 
Fig.\ref{mdot}. A good agreement is found between the estimated value of $\nu$ 
and that predicted for a non-relaxed system orbiting on a circular orbit.

\section{Discussion}

The evidence reported in the previous section strongly indicate that Palomar 14 
is currently in a stage of tidal disruption.

In fact, the cluster appears to extend up to 20'~ and beyond, i.e. at 
least 4 times the previous estimates of its tidal radius.
The decrease of the radial density profile indicates that the remnant has
a symmetry respect to the cluster center and cannot be addressed to any
foreground or background structure (see, e.g., Martinez-Delgado et al. 2002, Bellazzini
et al. 2003). In fact, the typical size of satellite system 
(such as the stream of an extinct galaxy) is $\sim$ 4 kpc (Mateo 1998) and should 
therefore be recognizable as a flat homogeneously distributed overdensity over 
the relatively small (1 sq. deg.) area covered by our observations. 

It is difficult to explain the anomalous distribution of cluster stars by simply
assuming a larger tidal radius. The stars located in the outermost region of the
cluster (at distances $r>$15'~ from the center) have indeed a very 
anisotropic distribution forming a collimated tail which is symmetrical on
both sides of the cluster. Moreover, the measured tidal radius exceeds the 
maximum Roche radius by more than a factor of two (see
Sect. \ref{struc}). 
In spite of the large uncertainties involved in the maximum Roche radius 
estimate, no stars are expected to be bound to the cluster at 
distances $r>10'$.    

The coherence and the symmetry of the overdensity measured in our sample
strongly suggest the presence of a pair of tidal tails surrounding this cluster. 
The ellipticity and the orientation of the tails appears to change as a function
of the distance from the cluster center. This is also predicted by dynamical
simulation of dissolving stellar systems in a Galactic tidal field (Montuori et
al. 2007; Klimentowsi et al. 2009). In fact, while the central region of the 
cluster is expected to be elongated towards the
Galactic center (see Fig. \ref{map}), the external regions should roughly be aligned with 
the cluster orbit. In the same way, while in the inner regions the tidal tails are expected to 
marginally affect the cluster ellipticity, in the outer regions they should
dominate the shape of the density contours increasing the measured ellipticity 
(see Fig. \ref{ang}).
Unfortunately, the
distance of this object prevents a direct measure of its proper motion.
A curious agreement is found between the direction of the tail and the
expected proper motion provided by Lynden-Bell \&
Lynden-Bell (1995) in the assumption that 
Palomar 14 belongs to a stream comprising Fornax and Palomar 15. 

This is the first time that a significant tidal tail is discovered in an outer
halo cluster ( at Galactocentric distances $R_{GC}>40~Kpc$). The tidal field in such an external region of the Galaxy is very small
and even a loose GC in a quasi-circular orbit is not expected to suffer a 
strong tidal stirring (Lee et al. 2006; Fellhauer \& Lin 2007). However, the
structure of Palomar 14 appears to be quite peculiar among Galactic GCs: it has
in fact the largest half-light radius and  a modest mass. Its low density
makes it very susceptible to the tidal strain of the Milky Way halo. This is also
confirmed by the location of the Roche radius inside the cluster tidal radius
and by the relatively high mass-loss rate (see Sect. \ref{struc}).

\begin{figure}
\epsscale{1.2}
\plotone{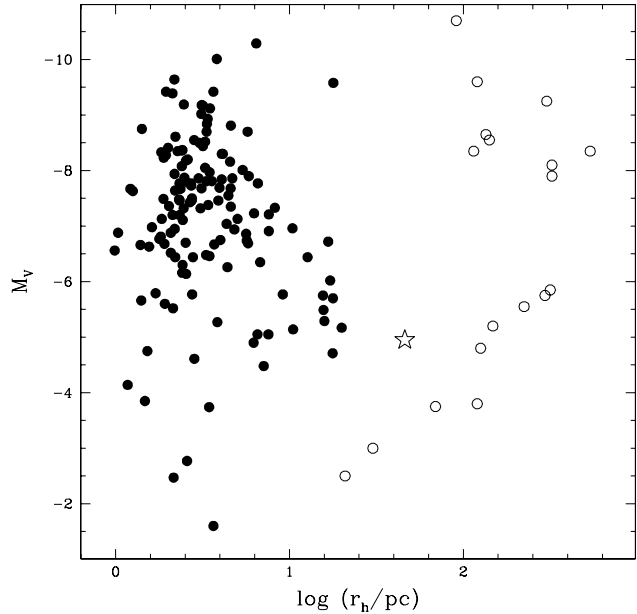}
\caption{Location of different classes of objects in the plane of absolute magnitude vs. 
half-light radius. Filled circles are Galactic GCs (from Harris 1996), open
circles are Milky Way satellites (Irwin \& Hatzidimitriou 1995; Mateo 1998; Willman et al. 2005;
Zucker et al. 2006a, 2006b; Belokurov et al. 2006b, 2007). The location 
of Palomar 14 is marked with an open star.}
\label{belok}
\end{figure}

The peculiar structural properties of Palomar 14 constrains its orbit and nature.  
In Fig. \ref{belok} the location of Palomar 14 in the $M_{V}-log~r_{h}$ diagram 
is shown along with Galactic GCs 
and satellites dwarf galaxies discovered over the past ten years.
Palomar 14 stands in an intermediate position between the 
loci of Galactic GCs and satellite galaxies.
The  area of the diagram populated by faint objects is a region 
where the separation between GCs and dwarf galaxies is perhaps somewhat blurred.  
Yet, evidence for  large diffuse clusters orbiting around \object{M31} (Chapman et al. 2008),
\object{NGC 1023} (Larsen \& Brodie 2000; Brodie \& Larsen 2002) and in the ACS Virgo 
Cluster survey (Peng et al. 2006) are now coming to light. 

We cannot exclude the possibility that the cluster is also the remnant of an ancient galaxy
progressively disrupted by the interaction with the Milky Way potential. In this case, 
however, the photometry and internal kinematics
severely constrain the scenario. First, the CMD presents
a narrow Red Giant Branch (RGB) and a
compact Red Clump (as found in previous HST studies; Dotter et al. 2008),
suggesting a very homogeneous chemical composition. Second, the recent spectroscopic analysis by Jordi et al. (2009) 
yields a very small velocity dispersion
($\sigma_{v}=0.38\pm0.12$km~s$^{-1}$) which translates into a
mass-to-light ratio $M/L<2.2~M_{\odot}/L_{\odot}$. Note that the M/L ratio of a 
galaxy remnant is expected to increase with time as a result of the
continuous loss of stars (Pe{\~n}arrubia et al. 2008).

The fraction of stars located in the tails indicates a significant
mass loss rate ($\nu>0.18~Gyr^{-1}$). An even larger rate is expected if
eccentric orbits are considered (see Sect. \ref{struc}). Consider that, for a
mass loss rate larger than $\nu>0.3~Gyr^{-1}$, a system with a typical GC mass
($M\sim 5 \cdot 10^{5} M_{\odot}$) would lose more than 95\% of its mass after 12
Gyr.
The key question is therefore how this old cluster has survived the tidal
interaction in spite of its small density. 
Two hypotheses, which are not mutually exclusive,  can be put forward: {\it (i)} 
Palomar 14 follows an external orbit confined to the
peripheral regions of the Galactic halo and/or {\it (ii)} it formed in a
satellite galaxy later accreted by the Milky Way.
In the former scenario, the cluster spent 
most of its evolution in a
peripheral region of the Galaxy (at distances always $>$60 kpc) and experienced
only a minor tidal stirring from the Milky Way which allowed its survival until
the present epoch.
In the latter scenario, the cluster evolved in an environment where it could
form and survive without suffering the strong tidal effect of the Milky Way. 

After all, in the case of M31, most, if not all, extended GCs at distances 
larger than 30 kpc appear to be associated with streams 
(Chapman et al. 2008; Forbes et al. 2010; Mackey et al. 2010). 
This seems to suggest that extended clusters follow different formation/evolutionary
processes depending on their birth places. It would therefore not be very surprising if Palomar 14 
would be yet another example of
the on-going accretion populating the stellar Galactic halo.

\acknowledgments

Based on observations obtained with MegaPrime/MegaCam,
a joint project of CFHT and CEA/DAPNIA, at the Canada-France-Hawaii Telescope 
(CFHT) which is operated by the National Research Council (NRC) of Canada, the 
Institut National des Science de l'Univers of the Centre National de la 
Recherche Scientifique (CNRS) of France, and the University of Hawaii.
This research was supported by the Spanish Ministry of Science and 
Innovation (MICINN) under grant AYA2007-65090, by 
CNRS/MAE-PICASSO and ANR POMMME(ANR 09-BLAN-0228). 
This work is based in part on data products produced at 
the TERAPIX data center located at the Institut d'Astrophysique de Paris.
We warmly thank Mark Gieles, the referee of our
paper, for the helpful comments and suggestions and Igor Drozdovsky for a
critical reading of the paper.\\

{\it Facilities:} \facility{CFHT}

\end{document}